\documentclass[a4paper]{article}
\usepackage{fixltx2e}
\usepackage{authblk}
\usepackage[latin1]{inputenc}
\usepackage{amsmath,amsfonts,amssymb,bm}
\usepackage{graphicx}
\usepackage{cite}
\usepackage[tight,hang,small,bf]{subfigure}

\begin{document}
\title{\textbf{Phase-stable source of polarization-entangled photons in a linear double-pass configuration}}
\author{Fabian Steinlechner,$^{1,*}$ Sven Ramelow,$^{2,3}$ Marc Jofre,$^1$ \\Marta Gilaberte,$^1$ Thomas Jennewein,$^4$ Juan. P. Torres,$^{1,5}$ \\Morgan W. Mitchell,$^{1,6}$ and Valerio Pruneri$^{1,6}$}
\affil{\small{
$^1$ICFO-Institut de Ciencies Fotoniques, 08860 Castelldefels (Barcelona), Spain \\
$^2$Institute for Quantum Optics and Quantum Information, Austrian Academy of Sciences,
Boltzmanngasse 3, 1090 Wien, Austria \\
$^3$ Quantum Optics, Quantum Nanophysics, Quantum Information, University of Vienna, Faculty of Physics, Boltzmanngasse 5, 1090 Wien, Austria \\
$^4$Institute for Quantum Computing and Department of Physics and Astronomy, University of Waterloo, Ontario N2L 3G1, Canada \\ 
$^5$Department of Signal Theory and Communications, Universitat Politecnica de Catalunya, Jordi Girona 1-3, 08034 Barcelona Spain\\
$^6$ICREA-Institució Catalana de Recerca i Estudis Avançats, 08010 Barcelona, Spain}}
\affil{*\underline{\emph{fabian.steinlechner@icfo.eu}}}
\date{}
\maketitle
\begin{abstract} 
We demonstrate a compact, robust, and highly efficient source of polarization-entangled photons, based on linear bi-directional down-conversion in a novel 'folded sandwich' configuration.  Bi-directionally pumping a single periodically poled KTiOPO$_4$ (ppKTP) crystal with a 405-nm laser diode, we generate entangled photon pairs at the non-degenerate wavelengths 784 nm (signal) and 839 nm (idler), and achieve an unprecedented detection rate of 11.8 kcps for 10.4 $\mu$W of pump power (1.1 million pairs / mW), in a 2.9-nm bandwidth, while maintaining a very high two-photon entanglement quality, with a Bell-state fidelity of $99.3\pm0.3$\%. 
\end{abstract}


\section{Introduction}\label{sec:Introduction}
Entangled photons are essential for many fundamental quantum optics experiments, as well as a key resource in quantum communication \cite{tittel} and the emerging field of quantum technologies \cite{Obrien:09}. Experiments in these fields are becoming increasingly demanding, and an efficient, high-quality source of entangled photons can now be considered a necessary tool in the quantum mechanic's toolbox. The quality of an entangled photon source is commonly characterized by its brightness, that is, the number of generated pairs per mW of pump power and per nm of generated bandwidth, as well as the purity of the  entangled state, or visibility. With experiments incorporating entangled photons expanding to the fields of biology and telecom engineering, other essential criteria are the ease of operation and long-term stability, as well as the number, cost and complexity of optical components required to build the source. 

To date the most widely used method for generating photon pairs is based on spontaneous parametric down conversion (SPDC) in bulk nonlinear crystals, such as potassium titanyl phosphate (KTP) and beta barium borate (BBO). Numerous schemes for polarization-entanglement of photon pairs generated in SPDC have been proposed, and demonstrated \cite{Kwiat:95,Kwiat:99,kim:01,Barbieri:03,Kuklewicz:04,fiorentino:04,shi:04,Kim:06}. In the case of  collinear SPDC, two commonly implemented configurations are based on crossed crystals and non-degenerate wavelengths \cite{Trojek:08}, or a bi-directionally pumped crystal in a Sagnac-loop \cite{shi:04,Kim:06}. These schemes share the advantage that no active interferometric stabilization is required, due to the common path configuration for down-converted and pump photons. Furthermore, using periodically poled (pp) crystals, they can be implemented in collinear, non-critically phase-matched configurations in materials with large nonlinear coefficients, which has led to some of the brightest, high-visibility sources of polarization-entangled photons demonstrated to date (see Table \ref{stateofart} for a comparison). Waveguide-based sources could lead to even higher efficiency \cite{Tanzilli:2012,Fiorentino:08}, but high brightness and polarization-entanglement-visibility are yet to be demonstrated simultaneously.
\begin{table}[h]
\caption{State-of-the-art sources of polarization-entangled photons, based on bulk SPDC in the near infrared region. The key performance characteristics are the detected pair-rate per mW of incident pump power (B), the full width at half maximum bandwidth (FWHM) of the generated photons ($\Delta \lambda$), the resulting spectral brightness ($B/\Delta \lambda$=B$^{(\lambda)}$) in pairs per mW per nm, and the Bell-state fidelity (F) achieved in the respective set-up. Ref. \cite{Predojevic:12,Rangarajan:09} correspond to pulsed sources. For a comprehensive comparison of sources until 2007, see Refs.\cite{Fedrizzi:07,Trojek:Phd}. To our knowledge, the highest total (un-normalized) pair-rate of 1.1 Mcps (F$\sim$0.97), was observed in \cite{Altepeter:05} for a pump power of 280 mW.}
\begin{center}
  \begin{tabular}{ | c | c |c |c | c | }
    \hline
	
    Reference & B [kcps/mW] & $\Delta \lambda$ [nm] & B$^{(\lambda)}$ [kcps/mW/nm] & F \\ \hline	\hline
		Fedrizzi et al.\cite{Fedrizzi:07} & 82 & 0.3 & 271 & 99.6 \% \\ 	
    Trojek et al.\cite{Trojek:08} & 27 & 14.5 & 1.8 & 99.4 \% \\ 
		Rangarajan et al.\cite{Rangarajan:09} & 13 & - & - & 99 \%\\
    Steinlechner et al.\cite{Steinlechner:12} & 640 & 2.3 & 278 & 98.3 \% \\ 
    Predojevi\'{c} et al.\cite{Predojevic:12} & 39.7 & 2.67 & 15 & 98.3 \%\\  

		\hline
		
  \end{tabular}\label{stateofart}
	
\end{center}
\end{table}
Here, we restrict our considerations to the bulk case, where the crossed-crystal configuration, also known as 'sandwich' configuration, is widely used due to its ease of alignment; an advantage that comes at the cost of a greater complexity associated with the use of two nonlinear crystals, in particular with long periodically poled crystals. When combining SPDC from two different down-converters it is crucial that the sources be identical, as any difference would result in incoherent summing over different phase-contributions, thus diminishing the purity of the entangled state. Due to technical issues in the crystal poling procedure, such differences are indeed observed \cite{Steinlechner:12}, and both efficiency and spectral characteristics of nonlinear crystals may vary - even within the same batch. This requires additional efforts, such as maintaining the two crystals at different phase-matching temperatures, or narrow-band spectral filtering (at the cost of efficiency) to erase detrimental which-crystal information, and achieve high entanglement visibility. The Sagnac scheme is not prone to such issues as it makes use of only a single down-converting crystal, but requires overlapping the SPDC emission in both arms of the interferometer in order to achieve high brightness and visibility. This involves an intricate alignment procedure, particularly time-consuming for less experienced users, which becomes even more daunting with the small beam waists required to achieve high pair-generation efficiencies.\\

Here we present a novel source of polarization-entangled photons which combines advantages of Sagnac and sandwich configurations, while using a limited number of optical components. The 'folded sandwich' scheme uses only one down-conversion crystal combined with a wave plate, to efficiently generate polarization-entangled photon pairs. In an experimental realization with an 11.48-mm-long ppKTP crystal we demonstrate both highly efficient pair-generation, as well as a high degree of polarization-entanglement over a large bandwidth, without the need for active interferometric stabilization. The source can easily be further integrated and is likely to become compliant with the severe requirements of space flight and operation.

\section{Linear double-pass scheme}\label{sec:principle}
\begin{figure}[htbp]
\centering\includegraphics[width=12cm]{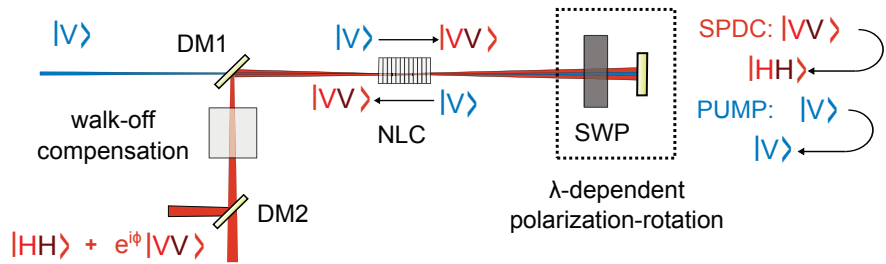}
\caption{The principle of operation can be understood as a 'sandwich' configuration, in which - instead of using a second orthogonally oriented crystal - the SPDC from a single nonlinear crystal (NLC) is transformed to the orthogonal polarization (i.e. 'folded sandwich'). }
\label{fig:scheme}
\end{figure}

We generate entanglement in a linear bi-directional down-conversion geometry as depicted in Fig. \ref{fig:scheme}.
A pair of collinearly propagating, co-polarized photons is generated via SPDC in a nonlinear crystal (e.g. ppKTP). Polarization-entanglement at the non-degenerate wavelengths can be achieved via a polarization-flipped round trip \cite{Hodelin:06,Barbieri:03,shi2:04}, where the photons emitted in the first pass (in state $\vert V_{\lambda_s} V_{\lambda_i} \rangle$) are transformed to $\vert H_{\lambda_s} H_{\lambda_i} \rangle$, via the double-pass through specifically tailored wave plate (SWP) oriented at 45$^\circ$, that introduces a quarter-wave retardation in the NIR, and a half- or full-wave retardation for the pump. The SPDC generated in the second-pass of the pump is emitted in the state $\vert V_{\lambda_s} V_{\lambda_i} \rangle$, and a polarization-entangled state
\begin{equation}
\vert \Psi(\phi) \rangle = \frac{1}{\sqrt{2}} \left(\vert V_{\lambda_s} V_{\lambda_i} \rangle+ e^{i \phi(\lambda_s,\lambda_i)} \vert H_{\lambda_s} H_{\lambda_i} \rangle\right)
\end{equation}
results if a fixed relative phase relationship $\phi(\lambda_s,\lambda_i)$ is maintained over the detected SPDC bandwidth. It is well-known that in SPDC, the summed phase of signal and idler photons equals the phase of the pump photons. The presented scheme thus does not require active path-length stabilization of the linear interferometer, since first-pass SPDC, and second-pass pump photons propagate along the same optical path \cite{Kim:06}. Hence path-length fluctuations are effectively auto-compensated, as reflected in the absence of the interferometer path-length in the relative phase in Eq. \ref{relphase}. We note that any spatial which-crystal information the photons may carry after SPDC emission is effectively erased via the projection into the single spatial mode of a coupling fiber (not depicted in Fig. \ref{fig:scheme}). Which-polarization information due to distinct spectral intensity profiles \cite{Steinlechner:12} is not present, as the scheme uses the same down-conversion crystal twice. However, distinguishing information remains in the spectral phase profile, and requires compensation - see section \ref{comp}.

The non-degenerate signal and idler photons are finally separated from the pump via a dichroic mirror (DM1), and transformed into two distinct spatial modes, using either a dichroic mirror (DM2) or fiber-based wavelength-division multiplexer. Note, that for the signal and idler to be efficiently isolated, this requires working at sufficiently non-degenerate wavelengths.
\subsection{Compensation crystals}\label{comp}
\begin{figure}[htbp]
\centering\includegraphics[width=6cm]{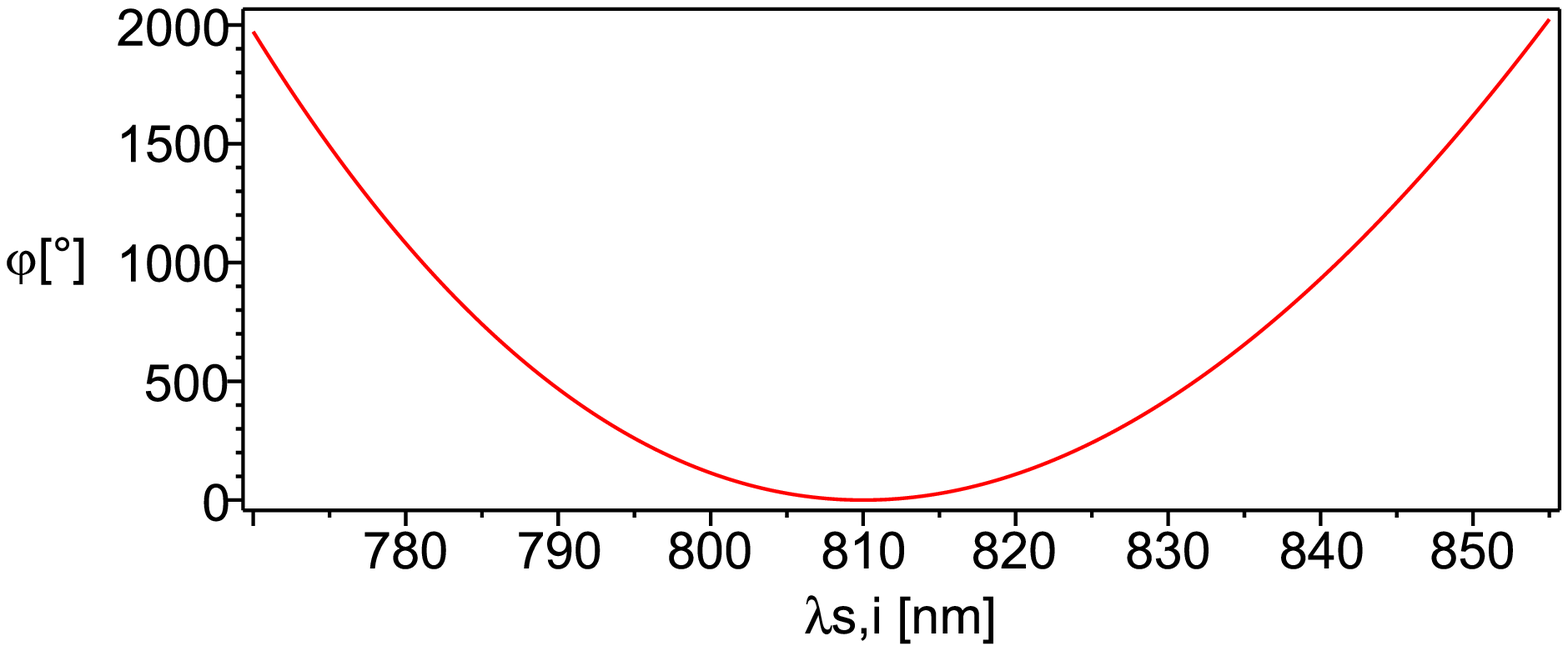}
\centering\includegraphics[width=6cm]{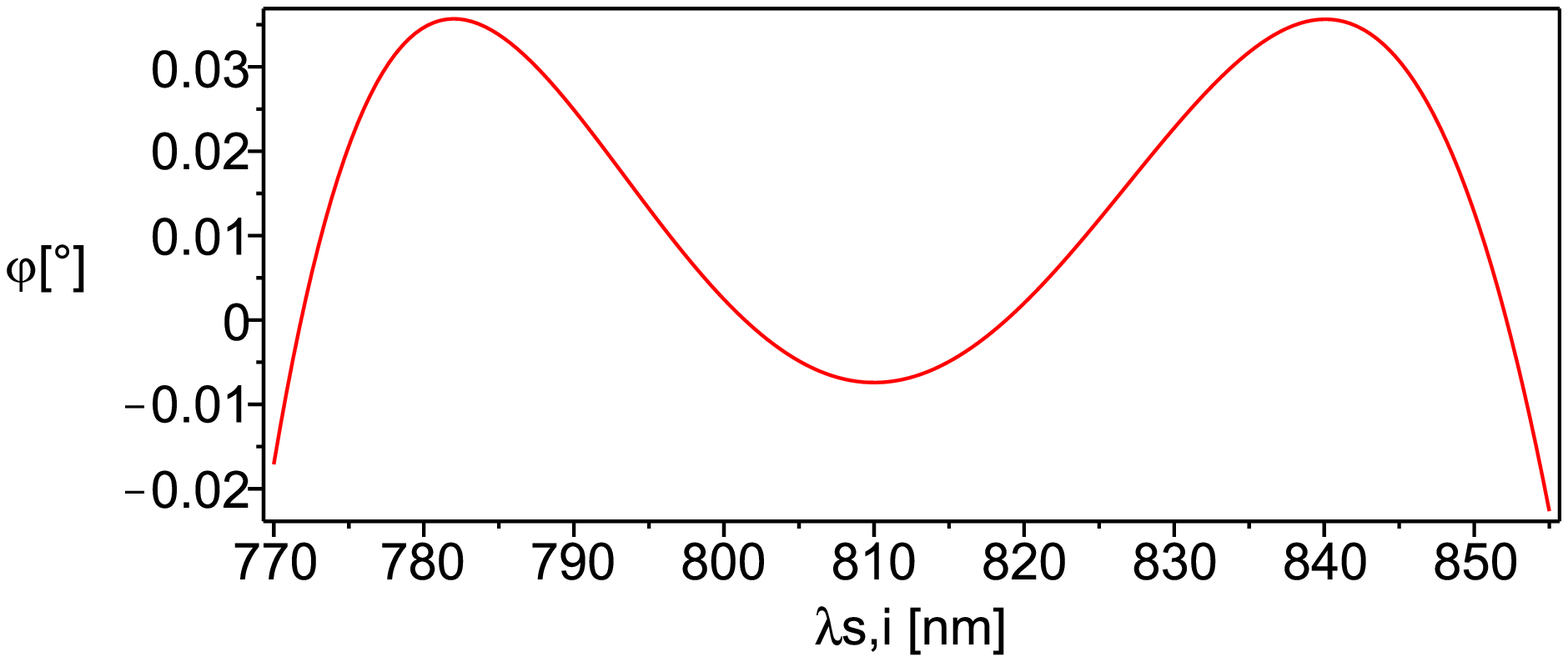}
\caption{(Left:) Strong wavelength-dependent phase variations (constant offset subtracted) between down-converted photons generated in the first- and second pass through the nonlinear crystal, due to chromatic dispersion in ppKTP and the wave plate. (Right:) Flattened phase variations over a broad spectral range after compensation with YVO$_4$ crystal. Note the difference in scale between both figures.}
\label{fig:phasemap}
\end{figure}

As a consequence of the double-pass through the dispersive nonlinear crystal, the  $\vert H_{\lambda_s} H_{\lambda_i} \rangle$  pairs generated in the first pass acquire different dispersive characteristics  than the  $\vert V_{\lambda_s} V_{\lambda_i} \rangle$ pairs emitted in the second pass. As this can yield which-crystal information, leading to the deterioration of the purity of the two-photon polarization state, compensation of this phase is required. For photons propagating along the crystallographic x-axis of a ppKTP crystal of length L, the accumulated phase difference, resulting from the phases acquired individually by horizontally and vertically polarized pump, signal, and idler photons ($\phi^{H,V}_{s,i,p}$), reads  
\begin{align} \label{relphase}
	\begin{split}
	\phi(\lambda_s,\lambda_i)						 			
				&=\phi^V_p+\phi^H_s+\phi^H_i-(\phi^V_s+\phi^V_i) \\
				&=\phi^V_p + 2 \pi L \left(\frac{n_y(\lambda_i)}{\lambda_i} 
+\frac{n_y(\lambda_s)}{\lambda_s}\right) +2 \phi_{qwp}(\lambda_s,\lambda_i).
	\end{split}					
\end{align}
For an 11.48-mm ppKTP crystal, and the double-pass through an achromatic quarter-wave plate made of MgF$_2$ and SiO$_2$, strong variations of the uncompensated relative phase appear, as depicted in Fig. \ref{fig:phasemap}. In order to counteract this effect a 18.5-mm YVO$_4$ crystal, exhibiting opposite birefringent characteristics is inserted after the down-converter. This way an 
\begin{equation}
\phi_{C}(\lambda_s,\lambda_i) = 2 \pi \times L_{YVO}\left[\frac{n^{(o)}(\lambda_s)}{\lambda_s}+\frac{n^{(o)}(\lambda_i)}{\lambda_i}-\left( \frac{n^{(e)}(\lambda_s)}{\lambda_s}+\frac{n^{(e)}(\lambda_i)}{\lambda_i} \right) \right]
\end{equation}
effectively flat phase can be maintained over a broad spectral range $\phi_{C}(\lambda_s,\lambda_i)+\phi(\lambda_s,\lambda_i)\approx C$. Note that here we neglect additional effects that would be present when using broadband pump \cite{Trojek:Phd,Nambu:02}. For long nonlinear crystals, and correspondingly scaled compensation crystals, the thermal dependence of the optical path-lengths of extra-ordinary and ordinary polarized photons, makes thermal isolation necessary in order to achieve good long-term stability. In the case of 18.5-mm of YVO$_4$, we calculate \cite{Zelmon:10} that a temperature change of $\sim 2.4^{\circ}$C results in a $\pi$ phase-shift, which is in good agreement with the experimental data (see section \ref{sec:setup}). This means, that in order to maintain a fidelity above 99.5 \% requires stabilization to $\pm$ 0.1\textdegree C. Similarly, temperature-dependent path-lengths in KTP \cite{Wiechmann:93,Emanueli:03} result in a required stability of the down-conversion crystal of $\pm$ 0.05\textdegree C.

\section{Experiment}\label{sec:setup}
\begin{figure}[htbp]
\centering
\includegraphics[width=12cm]{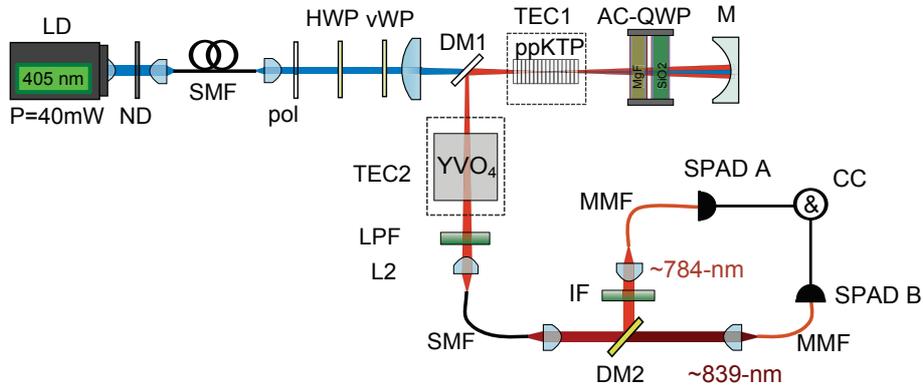}\label{setup-total}
\caption{Experimental realization of the 'folded sandwich' source of polarization-entangled photons. For details refer to main text.}
\end{figure}

\begin{figure}[htbp]
\centering\includegraphics[width=6cm]{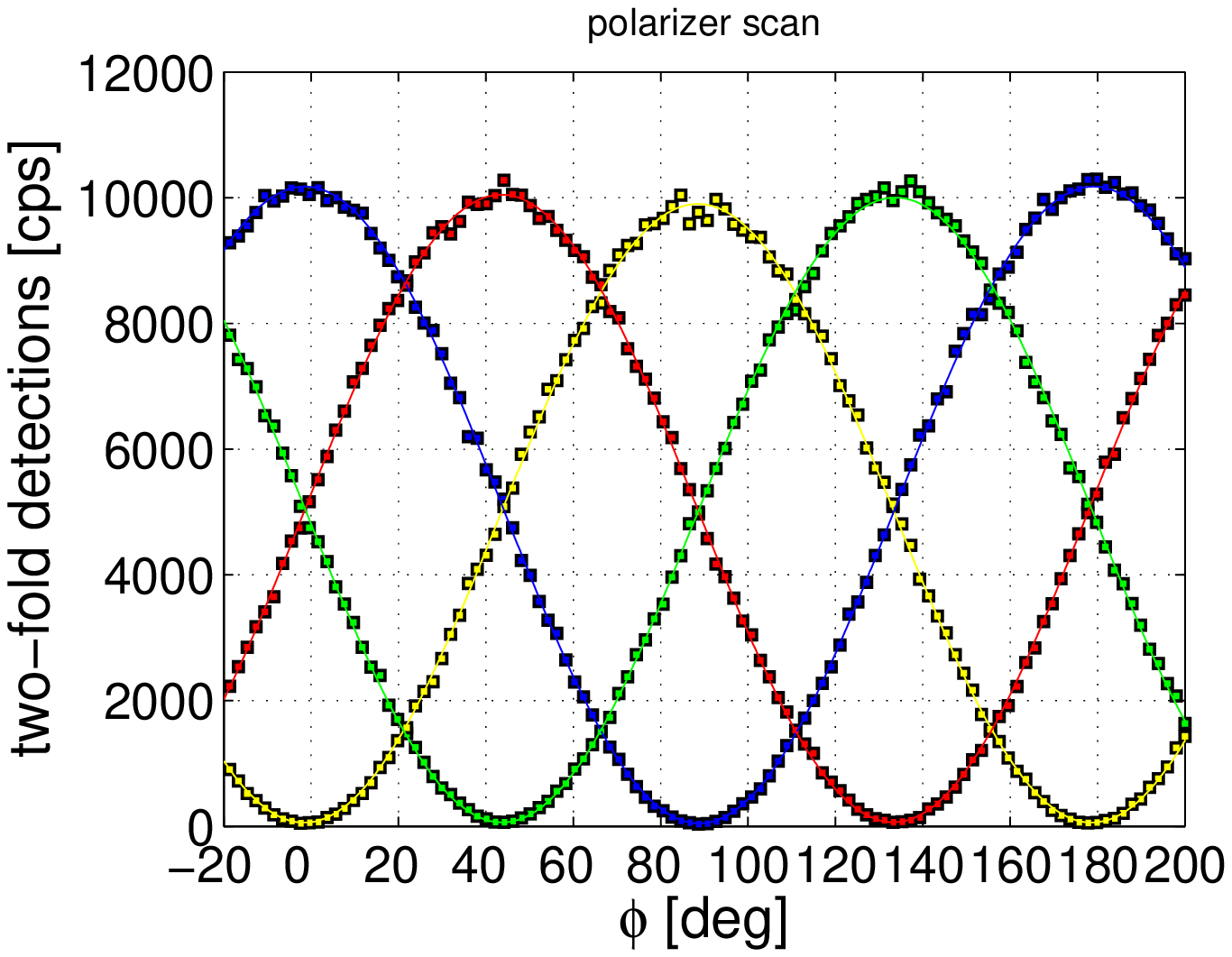}
\centering\includegraphics[width=6cm]{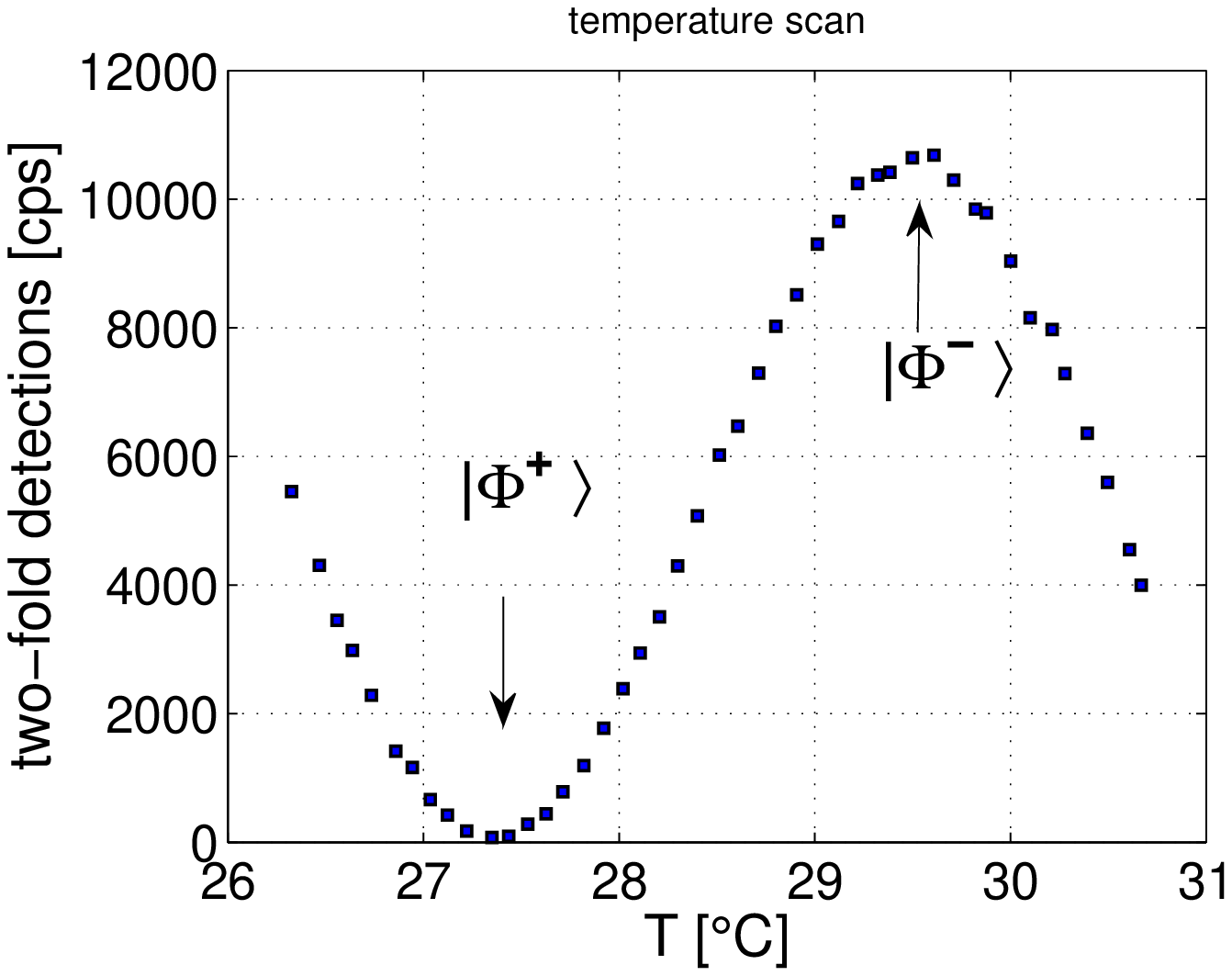}
\caption{(Left:) Coincidence counts per second, as a function of polarizer orientation $\theta_A$ in the signal arm, for $\theta_B$=$0^\circ$(blue), $90^\circ$(yellow), $45^\circ$(red), and $-45^\circ$(green). Square: experimental data, line: best fit. (Right:) Coincidence counts under angle-setting  $\theta_A$=45°,$\theta_B=-$45°, as the relative phase $\phi$ is modified by temperature-tuning YVO$_4$ compensation crystal.}
\label{fig:correlations}
\end{figure}

In the experimental setup depicted in Fig. \ref{setup-total}, a 11.48-mm-long ppKTP crystal (Raicol Crystals Ltd.), with a poling period of 3.425 $\mu$m, was mounted on an oven (TEC1), consisting of a Peltier element, a 10k$\Omega$ Thermistor, and a commercial temperature controller, with a specified long-term temperature stability of $\pm$0.002\textdegree C (Wavelength Electronics PTC2.5K-CH). The temperature was set to phase-match a collinear non-degenerate type-0 SPDC process, from 405.4 nm (pump) to  $\sim$784 nm (signal) and $\sim$839 nm (ilder). The fiber-coupled output beam of a volume holographic grating stabilized continuous wave (cw) laser diode (LD) (Ondax Inc.) with a center wavelength (cwl) of 405.4 nm was attenuated and focused to a beam waist of approximately 30 $\mu$m at the center of the nonlinear crystal. 
The polarization rotation of the $\vert V_{\lambda_s} V_{\lambda_i} \rangle$ photons to $\vert H_{\lambda_s} H_{\lambda_i} \rangle$ was implemented via the double-pass through a standard achromatic quarter-wave plate (AC-QWP), that was additionally anti-reflection coated at 405 nm and 750 nm - 850 nm, and oriented at 45$^\circ$ with respect to the ppKTP z-axis. For normal incidence the AC-QWP actually introduced a retardation of $\frac{0.93\lambda}{4}$ (tested at 850 nm and 785 nm), slightly different from the required $\frac{\lambda}{4}$ retardation. This was mitigated by a 3\textdegree \ tilt about the optical axis, under which it introduced the required retardation. At 405.4 nm, however, the AC-QWP introduced a retardation of $\frac{\lambda}{7}$ (as opposed to the full-$\lambda$ retardation, required for full usage of the pump-power in both passes), which lead to a reduction of the vertically polarized pump component in the second pass. This effect was pre-compensated by placing a birefringent phase- (vWP) and half-wave plates (HWP) in the pump path; setting azimuth and ellipticity of the pump photons accordingly, $\sim$80\% of the total pump power was polarized along the vertical ppKTP axis, and thus driving SPDC emission in both the first and second pass. The SPDC generated in both passes was spatially overlapped, locating the center of curvature of a broadband coated spherical mirror (M) at the crystal center $\sim z=R$, and extracted from the interferometer via a highly reflective dichroic mirror (DM1) that transmitted the pump photons. Dispersive de-phasing effects due to the double-pass were reversed by 18.3 mm of YVO$_4$ crystal. Additionally, the temperature-dependent birefringence of the YVO$_4$ crystal allowed the phase of the entangled state to be set via temperature-tuning (TEC2 - with same specifications as TEC1), as depicted in Fig. \ref{fig:correlations}. A color-glass long pass filter (LPF) blocked remaining pump and stray light, before the photons were coupled to single-mode fiber using an f=18.4-mm aspheric lens (L2) collecting a Gaussian beam-waist of $\sim 45\mu m$ located at the crystal center.

To assess the polarization correlations, the signal and idler photons were collimated and separated into two ports of free-space polarization analyzer, via a dichroic mirror with its transition edge angle tuned to 810 nm. The signal photons were then further filtered via an interference filter (IF) with a full width at half maximum (FWHM) pass-band of 3.5 nm and a peak transmission of $\sim$90\% (effectively filtering the idler-photons, in coincidence detection). After traversing the analyzer modules, the signal and idler were coupled into multi-mode fibers and guided to two single-photon avalanche diodes (SPAD) with a detection efficiency of $\sim$50\% ($\sim$300 cps dark counts), where coincident measurement events were recorded via a fast FPGA-based coincidence counting logic (CC), with the coincidence window set to 3.2 ns. In order to characterize the source independently from the detection system, the normalized pair-detection rate and entanglement were first assessed at low pump powers, at which accidental coincidences and SPAD saturation effects were negligible. With the IF placed in the signal path, we detected a total coincidence rate of $R_c$ = 11.8 kcps and singles rates of $R_s$ =61 kcps, and $R_i$ =88 kcps, at a pump power of 10.4 $\mu$W incident on the crystal. This  corresponds to an unprecedented \emph{detected} pair rate of 1.1 Mcps/mW and a detected spectral brightness of 0.39 Mcps/mW/nm (signal FWHM after IF $\sim$2.9 nm). We believe that pair-detection rates of several Mcps/mW are well within reach with this source design, and could be achieved by coupling the non-degenerate signal and idler photons to individual fibers, as well as using a longer nonlinear crystal and optimized beam waists \cite{Ljunggren:05,Bennink:10,Palacios:11}.\\

The polarization correlations were assessed by measurement of signal-idler coincidences in two mutually unbiased measurement bases (Fig. \ref{fig:correlations}) via motorized polarization analyzer modules (qutools quMotor), consisting of a quarter-wave plate (QWP) and a thin-film polarizer placed after DM2 (not depicted). The probability of measuring a signal-idler coincidence, as a function of the orientation of the two polarization analyzers ($\theta_A$ for the signal and $\theta_B$ for the idler photon) is $P(\theta_A,\theta_B)=\frac{1}{2}\left(1-Vsin(\theta_B-\theta_A)\right)$. Visibilities ($V$) were obtained as the best-fit parameter to the theoretically expected two-photon detection probability. Fixing $\theta_B$ to $0^\circ,90^\circ,45^\circ,$and $-45^\circ$, we measure raw visibilities of $V^{FIT}_{H}=98.8\%,V^{FIT}_{V}=99.1\%,V^{FIT}_{D}=98.7\%, V^{FIT}_{A}=98.8\%$, where the pump power at the crystal input facet was 0.028 mW in all cases. To further assess the quality of the entangled state, we measured the Bell-state fidelity via the fidelity witness  F=$\left\langle \Phi^{\pm} \left| \rho \right| \Phi^{\pm} \right\rangle= (1+V_{H/V} \pm V_{D/A} \mp V_{L/R})/4$ \cite{Guehne09}, where V$_{ij}=(N_{ii}+N_{jj}-N_{ij}-N_{ji})/(N_{ii}+N_{jj}+N_{ij}+N_{ji})$ are the visibilities in 3 mutually unbiased measurement bases (here ij=H/V,L/R,D/A), and $N_{ij}$ are the detected coincidence counts. At a pump power of 0.01 mW we obtain $V_{H/V}^{raw}=99.4\pm0.1 \%,V_{D/A}^{raw}=98.8\pm0.2\%,V_{L/R}^{raw}=98.7\pm0.2\%,$  and $V_{H/V}^{corr}=99.5\pm0.1\%,V_{D/A}^{corr}=99.0\pm0.2 \%,V_{L/R}^{corr}=98.9\pm0.2\%,$ before and after correction for accidental coincidence counts, respectively. This corresponds to  F$_{raw}=99.2 \pm 0.3\%$  and F$_{corr}=99.3\pm0.3\%$, confirming the high quality of polarization entanglement. 

\subsection{Scaling to high pump powers}
\begin{figure}[htbp]
\centering\includegraphics[width=6cm]{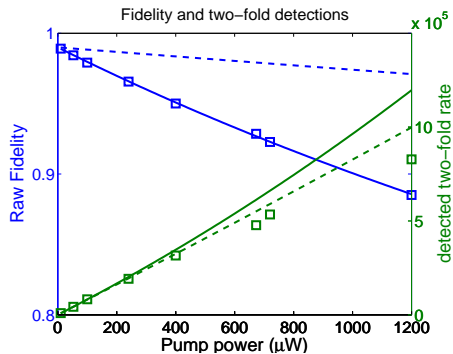}
\caption{Total rate of coincidences and raw state-fidelity, as a function of pump power. All experimental data (squares) was acquired with polarizers in place ($T\sim$0.9). The two-fold-rate was obtained by summing the coincidence-detections $C_{ij}$ over all polarizer positions in the H/V basis ($\sum_{ij=H/V}C_{ij}$). The simulations were conducted for a coincidence window of 3.2ns (solid lines), as used in the experimental setup , as well as 500ps (dashed lines).}

\label{two-fold-multipair}
\end{figure}

At higher pump powers, and correspondingly higher pair-rates, the limitations of the detection system employed lead to a decrease in fidelity and normalized detected pair-rate. For low count-rates, the scaling of experimental pair-rates is linear with the pump power. At higher pair-rates, the lower detection efficiency due to saturation effects leads to a decrease in two-fold rates. Figure \ref{two-fold-multipair} depicts the resulting deviation of observed two-fold rates from model calculations with ideal, saturation-free detectors (Simulated with the open access Matlab toolbox for quantum photonics \cite{Jenneweintoolbox}). Additionally, the timing resolution of the detection system (3.2ns), leads to an increased number uncorrelated photons being registered as coincidences  as the pair-rate increases (accidental coincidences), resulting in a lower raw fidelity for high pump powers. Note that these effects are however limitations of the detection system employed and could be mitigated utilizing an array of faster detectors. In our example an array of detectors with a timing resolution $\sim$500 ps \cite{Gallivanoni:06} would be required to directly measure the full 1.1Mcps at a fidelity above 97$\%$. For a more detailed analysis of the effect of multiple-pair detection, see e.g. Refs. \cite{Takesue:10,Jennewein:11,Steinlechner:12,Ramelow:13}.

\subsection{Verification of the phase-stability}\label{sec:stability}
\begin{figure}[htbp]
\centering\includegraphics[width=6cm]{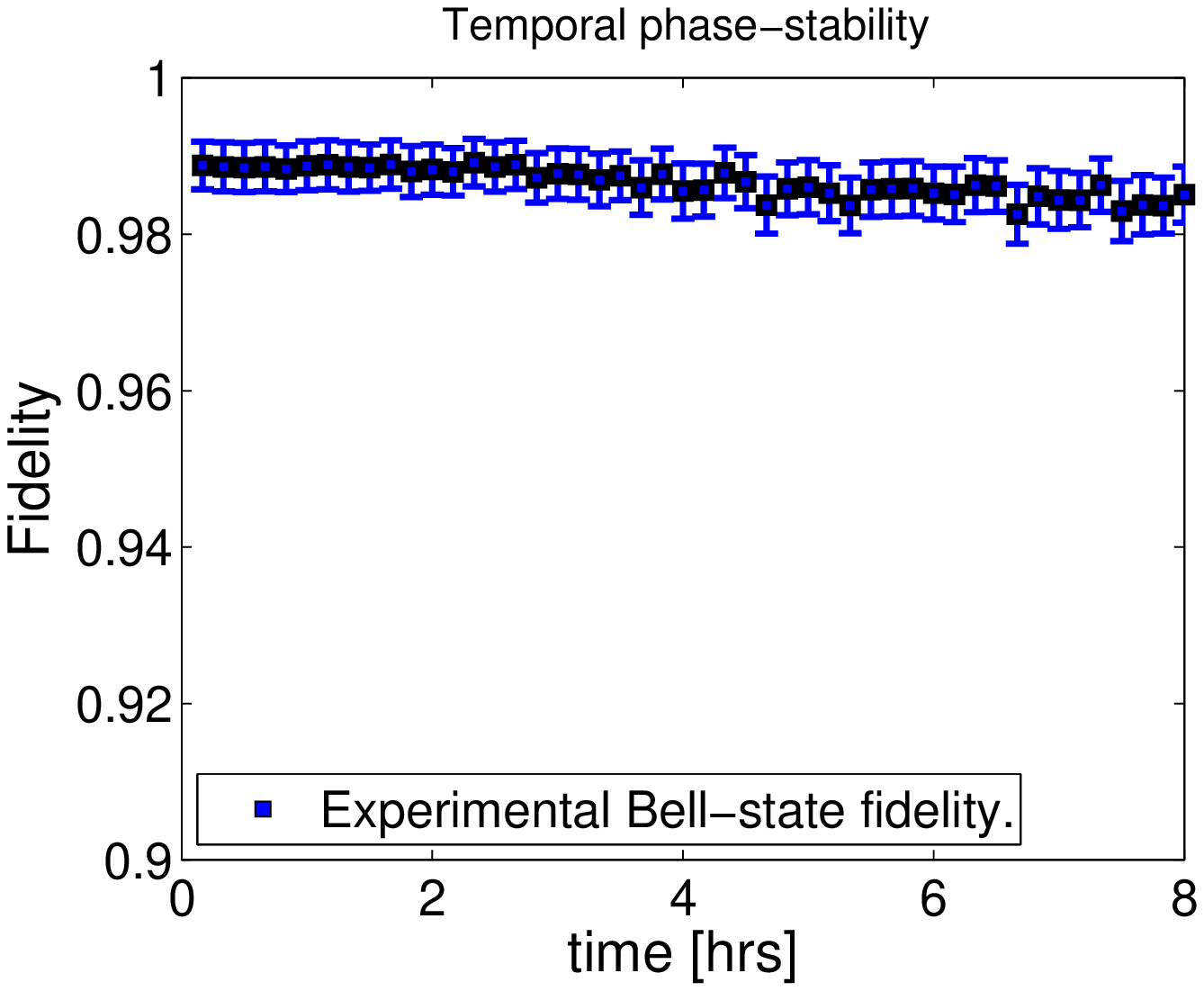}
\centering\includegraphics[width=6cm]{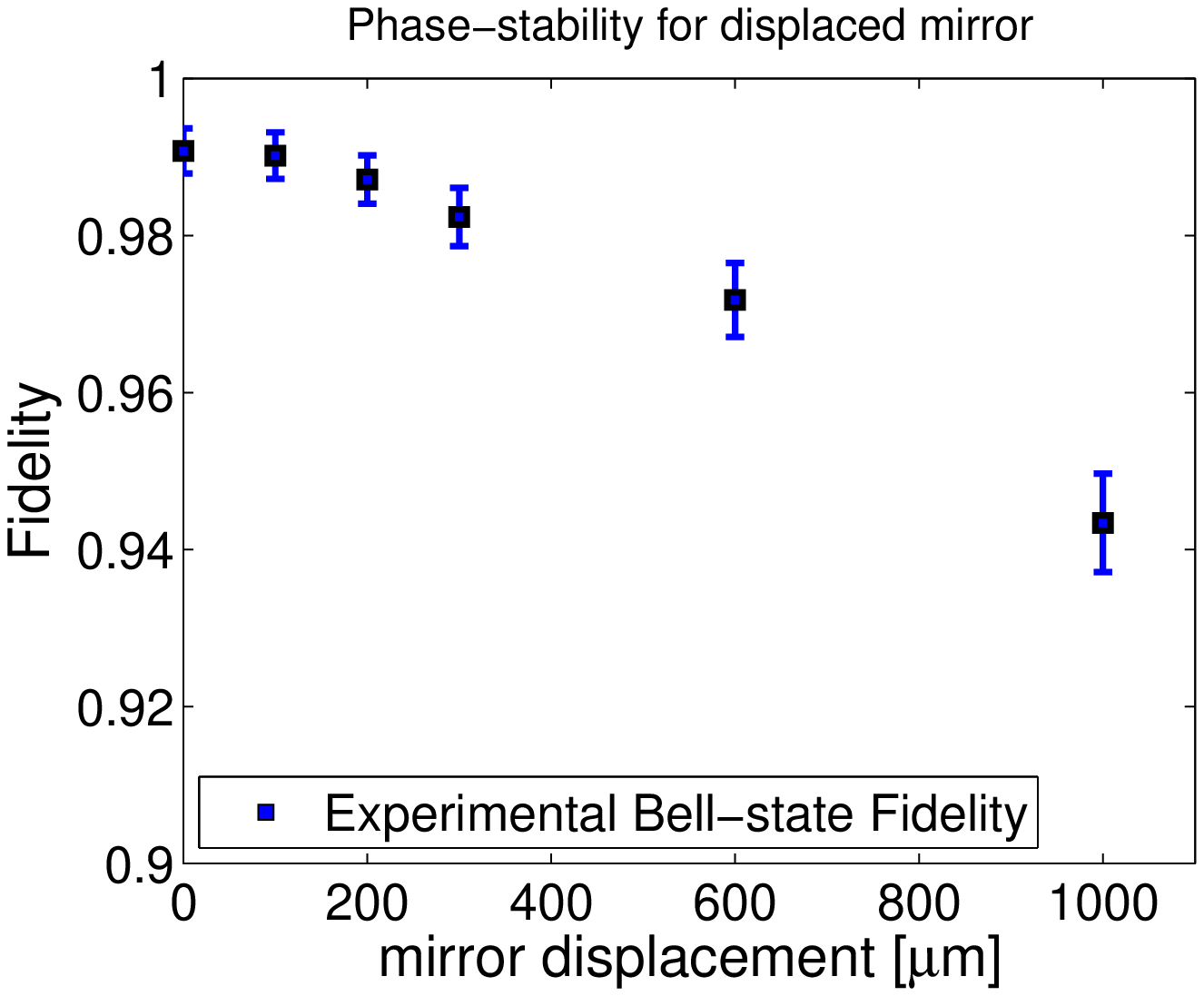}
\caption{(left:)  Bell-state fidelity over time, under laboratory conditions. (right:) Bell-state fidelity as the cavity length is increased, via displacement of mirror M.}
\label{stability}
\end{figure}

The phase-stability of the configuration was experimentally verified via measurements of the fidelity over time. The source showed good temporal stability, making long-term operation viable with periodic calibration on a time-scale of hours (see Fig. \ref{stability}). We believe that the stability could be even further improved, by placing a pre-compensation crystal in the pump path that effectively corrects for drifts of the pump cwl \cite{Nambu:02,Trojek:08}. To assess the mechanical stability requirements for the length of the double-pass cavity, we performed fidelity measurements for varied position of the retro-reflecting mirror (M). The result (Fig. \ref{stability}) shows that the phase remains constant for small displacements ($<$100$\mu$m). For larger displacements of the mirror, a phase-shift resulting from dispersion in air and geometric effects of the interacting Gaussian beams \cite{Predojevic:12}, as well as increasingly unbalanced fiber-coupling of first- and second-pass pair-emissions, lead to a decrease in fidelity with the initial state.  

\section{Conclusions}\label{sec:conclusions}
We have demonstrated a robust, compact, and highly efficient ($>$1 Mcps/mW) polarization-entangled photon source (Bell-state fidelity $>99\%$), based on linear bi-directional down-conversion in a folded sandwich configuration. The double-pass geometry eliminates the need for active interferometric stabilization, and allows the generation of extremely high-quality polarization entanglement via an effective temporal walk-off compensation scheme with a long YVO$_4$ crystal. Moreover, the compensation crystal also allows easy and precise control of the phase of the entangled state via temperature tuning. Comparing to previously reported sandwich and Sagnac schemes, the source allows recycling the first-pass pump power, and requires only a single nonlinear crystal. Furthermore, there is no efficiency trade-off when spatially overlapping SPDC emission in long crystals, which can be particularly useful in experiments where high heralding-efficiency is critical. The high demonstrated efficiency and entanglement visibility, combined with ease of alignment and phase stability make the source a valuable addition to existing schemes, that could find immediate application in the implementation of long-distance free-space quantum communications and cryptography links, as well as in quantum optics laboratory experiments. The use of a limited number of optical elements makes the source extremely well-suited for further integration and an ideal candidate for future applications in space.
\section*{Acknowledgments}
The authors thank Pavel Trojek for his contribution in early stages of design, as well as helpful discussion on the phase-compensation scheme. FS acknowledges financial support via FPI fellowship of the Spanish Ministry of Science and Innovation (MICINN). SR supported by the ERC (Advanced Grant QIT4QAD, 227844). JPT acknowledges support from the project PHORBITECH (FET-Open grant number 255914). Project funding via contracts TEC2010-14832, FIS2010-14831 and FIS2011-23520 funded by the Spanish MICINN. 

\bibliographystyle{osajnl}

\begin{thebibliography}{10}
\newcommand{\enquote}[1]{``#1''}

\bibitem{tittel}
W.~Tittel and G.~Weihs, \enquote{Photonic entanglement for fundamental tests
  and quantum communication,} Quantum Inf. and Comp. \textbf{1}, 3--56 (2001).

\bibitem{Obrien:09}
J.~L. O'Brien, A.~Furusawa, and J.~Vucovic, \enquote{Photonic quantum
  technologies,} Nat. Photonics \textbf{3}, 687--695 (2009).

\bibitem{Kwiat:95}
P.~G. Kwiat, K.~Mattle, H.~Weinfurter, A.~Zeilinger, A.~V. Sergienko, and
  Y.~Shih, \enquote{New high-intensity source of polarization-entangled photon
  pairs,} Phys. Rev. Lett. \textbf{75}, 4337--4341 (1995).

\bibitem{Kwiat:99}
P.~G. Kwiat, E.~Waks, A.~G. White, I.~Appelbaum, and P.~H. Eberhard,
  \enquote{Ultrabright source of polarization-entangled photons,} Phys. Rev. A
  \textbf{60}, R773--R776 (1999).

\bibitem{kim:01}
Y.-H. Kim, M.~V. Chekhova, S.~P. Kulik, M.~H. Rubin, and Y.~Shih,
  \enquote{Interferometric bell-state preparation using
  femtosecond-pulse-pumped spontaneous parametric down-conversion,} Phys. Rev.
  A \textbf{63}, 062301 (2001).

\bibitem{Barbieri:03}
M.~Barbieri, F.~De~Martini, G.~Di~Nepi, P.~Mataloni, G.~M. D'Ariano, and
  C.~Macchiavello, \enquote{Detection of entanglement with polarized photons:
  Experimental realization of an entanglement witness,} Phys. Rev. Lett.
  \textbf{91}, 227901 (2003).

\bibitem{Kuklewicz:04}
C.~E. Kuklewicz, M.~Fiorentino, G.~Messin, F.~N.~C. Wong, and J.~H. Shapiro,
  \enquote{High-flux source of polarization-entangled photons from a
  periodically poled ${\mathrm{ktiopo}}_{4}$ parametric down-converter,} Phys.
  Rev. A \textbf{69}, 013807 (2004).

\bibitem{fiorentino:04}
M.~Fiorentino, G.~Messin, C.~E. Kuklewicz, F.~N.~C. Wong, and J.~H. Shapiro,
  \enquote{Generation of ultrabright tunable polarization entanglement without
  spatial, spectral, or temporal constraints,} Phys. Rev. A \textbf{69}, 041801
  (2004).

\bibitem{shi:04}
B.-S. Shi and A.~Tomita, \enquote{Generation of a pulsed polarization entangled
  photon pair using a sagnac interferometer,} Phys. Rev. A \textbf{69}, 013803
  (2004).

\bibitem{Kim:06}
T.~Kim, M.~Fiorentino, and F.~N.~C. Wong, \enquote{Phase-stable source of
  polarization-entangled photons using a polarization sagnac interferometer,}
  Phys. Rev. A \textbf{73}, 012316 (2006).

\bibitem{Trojek:08}
P.~Trojek and H.~Weinfurter, \enquote{Collinear source of
  polarization-entangled photon pairs at nondegenerate wavelengths,} Appl.
  Phys. Lett. \textbf{92}, 211103 (2008).

\bibitem{Tanzilli:2012}
S.~Tanzilli, A.~Martin, F.~Kaiser, M.~De~Micheli, O.~Alibart, and D.~Ostrowsky,
  \enquote{On the genesis and evolution of integrated quantum optics,} Laser
  Photon. Rev. \textbf{6}, 115--143 (2012).

\bibitem{Fiorentino:08}
M.~Fiorentino and R.~G. Beausoleil, \enquote{Compact sources of
  polarization-entangled photons,} Opt. Express \textbf{16}, 20149--20156
  (2008).

\bibitem{Predojevic:12}
A.~Predojevi\'{c}, S.~Grabher, and G.~Weihs, \enquote{Pulsed sagnac source of
  polarization entangled photon pairs,} Opt. Express \textbf{20}, 25022--25029
  (2012).

\bibitem{Rangarajan:09}
R.~Rangarajan, M.~Goggin, and P.~Kwiat, \enquote{Optimizing type-i
  polarization-entangled photons,} Opt. Express \textbf{17}, 18920--18933
  (2009).

\bibitem{Fedrizzi:07}
A.~Fedrizzi, T.~Herbst, A.~Poppe, T.~Jennewein, and A.~Zeilinger, \enquote{A
  wavelength-tunable fiber-coupled source of narrowband entangled photons,}
  Opt. Express \textbf{15}, 15377--15386 (2007).

\bibitem{Trojek:Phd}
P.~Trojek, \enquote{Efficient generation of photonic entanglement and
  multiparty quantum communication,} Ph.D. thesis, LMU-Munich (2007).

\bibitem{Altepeter:05}
J.~Altepeter, E.~Jeffrey, and P.~Kwiat, \enquote{Phase-compensated ultra-bright
  source of entangled photons,} Opt. Express \textbf{13}, 8951--8959 (2005).

\bibitem{Steinlechner:12}
F.~Steinlechner, P.~Trojek, M.~Jofre, H.~Weier, D.~Perez, T.~Jennewein,
  R.~Ursin, J.~Rarity, M.~W. Mitchell, J.~P. Torres, H.~Weinfurter, and
  V.~Pruneri, \enquote{A high-brightness source of polarization-entangled
  photons optimized for applications in free space,} Opt. Express \textbf{20},
  9640--9649 (2012).

\bibitem{Hodelin:06}
J.~F. Hodelin, G.~Khoury, and D.~Bouwmeester, \enquote{Optimal generation of
  pulsed entangled photon pairs,} Phys. Rev. A \textbf{74}, 013802 (2006).

\bibitem{shi2:04}
B.-S. Shi and A.~Tomita, \enquote{Preparation of a pulsed polarization
  entangled photon pair via interference,} Optics Communications \textbf{235},
  247 -- 252 (2004).

\bibitem{Nambu:02}
Y.~Nambu, K.~Usami, Y.~Tsuda, K.~Matsumoto, and K.~Nakamura,
  \enquote{Generation of polarization-entangled photon pairs in a cascade of
  two type-i crystals pumped by femtosecond pulses,} Phys. Rev. A \textbf{66},
  033816 (2002).

\bibitem{Zelmon:10}
D.~E. Zelmon, J.~J. Lee, K.~M. Currin, J.~M. Northridge, and D.~Perlov,
  \enquote{Revisiting the optical properties of nd doped yttrium
  orthovanadate,} Appl. Opt. \textbf{49}, 644--647 (2010).

\bibitem{Wiechmann:93}
W.~Wiechmann, S.~Kubota, T.~Fukui, and H.~Masuda, \enquote{Refractive-index
  temperature derivatives of potassium titanyl phosphate,} Opt. Lett.
  \textbf{18}, 1208--1210 (1993).

\bibitem{Emanueli:03}
S.~Emanueli and A.~Arie, \enquote{Temperature-dependent dispersion equations
  for ktiopo4 and ktioaso4,} Appl. Opt. \textbf{42}, 6661--6665 (2003).

\bibitem{Ljunggren:05}
D.~Ljunggren and M.~Tengner, \enquote{Optimal focusing for maximal collection
  of entangled narrow-band photon pairs into single-mode fibers,} Phys. Rev. A
  \textbf{72}, 062301 (2005).

\bibitem{Bennink:10}
R.~S. Bennink, \enquote{Optimal collinear gaussian beams for spontaneous
  parametric down-conversion,} Phys. Rev. A \textbf{81}, 053805 (2010).

\bibitem{Palacios:11}
S.~Palacios, R.~de~J.~Le\'{o}n-Montiel, M.~Hendrych, A.~Valencia, and J.~P.
  Torres, \enquote{Flux enhancement of photons entangled in orbital angular
  momentum,} Opt. Express \textbf{19}, 14108--14120 (2011).

\bibitem{Guehne09}
O.~Gühne and G.~Toth, \enquote{Entanglement detection,} Phys. Rep.
  \textbf{474}, 1 -- 75 (2009).

\bibitem{Jenneweintoolbox}
T.~Jennewein, \enquote{T. toolbox for quantum photonics in matlab.
  http://info.iqc.ca/qpl/ (accessed june 1, 2010).} .

\bibitem{Gallivanoni:06}
A.~Gallivanoni, I.~Rech, D.~Resnati, M.~Ghioni, and S.~Cova,
  \enquote{Monolithic active quenching and picosecond timing circuit suitable
  for large-area single-photon avalanche diodes,} Opt. Express \textbf{14},
  5021--5030 (2006).

\bibitem{Takesue:10}
H.~Takesue and K.~Shimizu, \enquote{Effects of multiple pairs on visibility
  measurements of entangled photons generated by spontaneous parametric
  processes,} Opt. Commun. \textbf{283}, 276--287 (2010).

\bibitem{Jennewein:11}
T.~Jennewein, M.~Barbieri, and A.~G. White, \enquote{Single-photon device
  requirements for operating linear optics quantum computing outside the
  post-selection basis,} J. Mod. Opt. \textbf{58}, 276--287 (2011).

\bibitem{Ramelow:13}
S.~Ramelow, A.~Mech, M.~Giustina, S.~Gr\"{o}blacher, W.~Wieczorek, J.~Beyer,
  A.~Lita, B.~Calkins, T.~Gerrits, S.~W. Nam, A.~Zeilinger, and R.~Ursin,
  \enquote{Highly efficient heralding of entangled single photons,} Opt.
  Express \textbf{21}, 6707--6717 (2013).

\end{thebibliography}

\end{document}